\documentclass[runningheads]{llncs}

\usepackage{graphicx}
\usepackage[hyphens]{url}
\usepackage{comment}
\usepackage[dvipsnames]{xcolor}
\definecolor{matrixBlue}{RGB}{102, 182, 222}
\definecolor{matrixYellow}{RGB}{246, 180, 81}
\definecolor{matrixGreen}{RGB}{60, 132, 116}

\usepackage{graphicx}
\usepackage{xcolor}

\usepackage{calc}

\newlength{\DepthReference}
\newlength{\HeightReference}
\newlength{\Width}%

\newcommand{\MyColorBox}[2][red]%
{%
    \settowidth{\Width}{#2}%
    \colorbox{#1}%
    {%
        \raisebox{-\DepthReference}%
        {%
                \parbox[b][\HeightReference+\DepthReference][c]{\Width}{\centering#2}%
        }%
    }%
}

\newcommand{\matrixElement}[2]{\colorbox{#1}{\textcolor{white}{\textsf{\textbf{#2}}}}}

\usepackage{csquotes}
\usepackage{todonotes}
\usepackage{cite}
\usepackage{float}
\usepackage{wrapfig}
\usepackage{picins}

\usepackage{scalerel}
\usepackage{tikz}
\usetikzlibrary{svg.path}
\definecolor{orcidlogocol}{HTML}{A6CE39}
\tikzset{
  orcidlogo/.pic={
    \fill[orcidlogocol] svg{M256,128c0,70.7-57.3,128-128,128C57.3,256,0,198.7,0,128C0,57.3,57.3,0,128,0C198.7,0,256,57.3,256,128z};
    \fill[white] svg{M86.3,186.2H70.9V79.1h15.4v48.4V186.2z}
                 svg{M108.9,79.1h41.6c39.6,0,57,28.3,57,53.6c0,27.5-21.5,53.6-56.8,53.6h-41.8V79.1z M124.3,172.4h24.5c34.9,0,42.9-26.5,42.9-39.7c0-21.5-13.7-39.7-43.7-39.7h-23.7V172.4z}
                 svg{M88.7,56.8c0,5.5-4.5,10.1-10.1,10.1c-5.6,0-10.1-4.6-10.1-10.1c0-5.6,4.5-10.1,10.1-10.1C84.2,46.7,88.7,51.3,88.7,56.8z};
  }
}
\newcommand\orcidicon[1]{\href{https://orcid.org/#1}{\mbox{\scalerel*{
\begin{tikzpicture}[yscale=-1,transform shape]
\pic{orcidlogo};
\end{tikzpicture}
}{|}}}}

\usepackage[colorlinks=true, allcolors=blue]{hyperref}
\begin{document}
\title{Cloud Native Privacy Engineering\\ through DevPrivOps}

\author{Elias Grünewald\orcidID{0000-0001-9076-9240}}

\authorrunning{E. Grünewald}
\institute{Technische Universität Berlin\\
Information Systems Engineering\\
Berlin, Germany\\
\email{gruenewald@tu-berlin.de}
}

\maketitle              %

\begin{abstract}
Cloud native information systems engineering enables scalable and resilient software architectures powering major online offerings. Today, these are built following agile development practices. At the same time, a growing demand for privacy-friendly services is articulated by societal norms and policy through effective legislative frameworks. In this paper, we \textit{(i)} identify conceptual dimensions of cloud native privacy engineering -- that is, bringing together cloud computing fundamentals and privacy regulation -- and propose an integrative approach to be addressed \textit{(a)} to overcome the shortcomings of existing privacy enhancing technologies in practice and \textit{(b)} evaluating existing system designs. Furthermore, we \textit{(ii)} propose a reference software development lifecycle called DevPrivOps to enhance established agile development methods with respect to privacy. Altogether, we show that cloud native privacy engineering opens up key advances to the state of the art of privacy by design and by default using latest technologies. 
\end{abstract}


\section{Introduction}
The enormous and unstoppable rise of digital services for people's lives already resulted in globally interconnected digital societies. During this long-lasting process %
the inter- and trans-disciplinary questions on how to achieve an adequate level of privacy are still to be solved -- while privacy itself is an essentially contested concept \cite{mulligan2016privacy}. Although some seem to have accepted sheer insurmountable hurdles or are actively supporting a post-privacy age (as shown by \cite{rauhofer}), many others, fortunately, fight for autonomy and against a \enquote{surveillance capitalism} \cite{zuboff2019age}; may it be through political advocacy, privacy law, or key technological advances. In this paper, we mainly focus on the latter with respect to current trends in the field of privacy engineering.

\pagebreak

All major digital service offerings are enabled through the extensive use of highly distributed cloud computing systems. These provision compute, storage, and network resources, that are used to build and run scalable and dynamic infrastructure and applications \cite{cncf}. Within the last decade, the service portfolio of public cloud vendors has bloomed from distributed databases over service meshes to highly-specific AI-based programming and execution platforms. However, not only the technical infrastructure has drastically changed, but also development models to create and operate distributed services. Software is crafted by diverse teams in agile programming, testing and design prototyping phases, and through iterative requirements engineering and using project management tools. Namely, agile development processes like scrum allow to develop and deploy new functionalities and complete services to production continuously (DevOps), i.e. potentially multiple times per hour \cite{bass2015devops}.

Inherently, distributed services are highly complex, which is why software engineering increasingly focuses on manageability, resilience and robustness, or observability to cope with the engineering challenges and -- as a secondary concern -- legal obligations of privacy and cloud computing. At the same time, the still emerging field of privacy engineering \cite{gurses2016privacy} has to provide the most accessible conceptual methods and technical tools to achieve privacy by design and by default, as legally required by the European General Data Protection Regulation (GDPR) \cite{gdpr} and commonly agreed upon in privacy research. Although, fundamental privacy principles \cite{cavoukian2009privacy} such as transparency, purpose limitation, and accountability, have been long established and are more often enforced \cite{Voigt2017}, so far, many developers lack a solid understanding and the concrete technologies to construct privacy-friendly cloud native systems. In short, we observe three major challenges:

\begin{itemize}
    \item \textbf{Cloud native application architectures introduce new privacy challenges} w.r.t. distributed (personal) data management across countries, availability under immense loads, compliant information flow control, restrictive access policies et cetera.
    \item \textbf{Software engineers are ill-equipped with privacy-preserving methods and tools addressing \textit{all} privacy principles}, including, among others, lawfulness, transparency, or accountability; while privacy is often misinterpreted as only subject to security-related research.
    \item \textbf{Agile development practices still (mostly) neglect or even contradict privacy principles} (beyond data minimization and security) as cross-cutting themes of software engineering.
\end{itemize}

\noindent Addressing these issues well aligns with related work on privacy and (early) cloud computing \cite{zhou}, %
engineering privacy by design \cite{bednar, cavoukian2020}, %
and, how privacy is affected by agile development practices \cite{gurses2018agile}. In a similar vein, this paper aims to provide a more clear viewpoint on the term of cloud native privacy engineering through a two-fold contribution:

\begin{itemize}
    \item A conceptual model on the \textbf{dimensions of cloud native privacy engineering} accompanied by different use case scenarios from an information systems engineering perspective, and
    \item Proposing a \textbf{privacy-aware \textit{DevPrivOps} reference lifecycle} addressing the shortcomings of established agile practices explicitly tailored to cloud native environments.  
\end{itemize}

\noindent The journey through this paper %
takes place as follows: First, we briefly introduce the established concepts of cloud native application architectures and agile software development and, further, compare to related work in sec.~\ref{related-work}.
On this basis, we observe the dimensions of cloud native privacy engineering in sec.~\ref{dimensions-cloud-native}
illustrated by several use case scenarios. Afterwards, we introduce the software development cycle called \textit{DevPrivOps} proposed for privacy-aware information systems engineering in sec.~\ref{dev-priv-ops}.
Finally, we discuss our findings and conclude in sec.~\ref{discussion}.

\section{Background and Related Work} \label{related-work}
This section introduces a brief background on the field of cloud native engineering and agile software development. Moreover, we summarize the latest findings in the field of privacy engineering.

\subsection{Cloud Native \& Agile Software Development}

Within the last decade, the technical evolution of distributed service-oriented architectures has been rapid and disruptive \cite{nieuwenhuis2018shift}. The emergence of cloud computing, mainly characterized by on-demand access to shared compute, storage, and network resources \cite{mell2011nist, marston2011cloud}, has led to a diverse and powerful infrastructure, platform, and software service portfolio \cite{lenk2009s, srivastava2018review}. Without doubt, the transformative power of cloud-based systems serves as an important utility across many dimensions of today's societies \cite{gill2019transformative}. Most prominently, major public cloud vendors such as Amazon Web Services, the Google Cloud Platform, Microsoft Azure, and IBM Cloud, showcase their offerings, which are adopted by a multitude of private and governmental customers. Furthermore, private and hybrid cloud approaches also enable online services. The latter are often powered by open source projects such as OpenStack\footnote{See \url{https://www.openstack.org/}.}.

To build and operate applications, which are scalable for millions of users, developers rely on so-called cloud native technologies. The Cloud Native Computing Foundation (CNCF) highlights the usage of \enquote{containers, service meshes, microservices, immutable infrastructure, and declarative APIs} \cite{cncf}. In practice, modern applications may consist of hundreds of loosely-coupled microservices that communicate through well-defined programming interfaces following paradigms such as REST \cite{dragoni2017microservices} or (g)RPC\footnote{See, e.g.,  \url{https://developers.googleblog.com/2015/02/introducing-grpc-new-open-source-http2.html}.}. 

At the same time, we observe a transformation from a (often waterfall-like) legacy software development culture towards a more flexible, iterative and agile organizational setup \cite{rajkumar2016devops, bass2015devops}. DevOps is widely acknowledged as the best way to deal with the complexity of large microservice architectures \cite{erich2017qualitative}. In doing so, the development team should be responsible for the entire lifecycle (incl. plan, code, build, test, release, deploy, operate, and monitor phases) of a software component and their expertise may be used to make individual technology decisions \cite{kratzke2017understanding}. Together with an adhered framework for managing tasks and responsibilities (such as scrum \cite{schwaber2004agile}), which integrates reasonable tool support for assisting all phases, fast-paced development with which high quality software can be achieved.

Finally, cloud native architecture, engineering, and management techniques heavily focus on the possible trade-offs between different software qualities and, moreover, ultimate technology decisions \cite{gannon2017cloud, balalaie2016microservices}. Such trade-offs occur in different shapes and sizes. They vary from evidence-based benchmarking experiments for choosing a best-fit technology to multilateral discussions on, e.g., what an adequate level of fair computing practice actually is in the context of cloud-based systems \cite{tai2016continuous}.

\subsection{Privacy} %

Privacy is a fundamental human right according to Art. 12 of the Universal Declaration of Human Rights \cite{un1948udhr}. Moreover, it has an even longer tradition as a societal norm and guideline for legislation and jurisdiction \cite{theRightToPrivacy}. Consequently, it is subject to inter- and trans-disciplinary research with legal, social, economic, political, psychological, and technical discourse. Predominantly, the notion of privacy is shaped by two different western cultures \cite{whitman2003two}. Being well aware of the different interpretations of privacy and data protection (including informational self-determination), hereafter we use these terms interchangeably. 

Today, privacy law (and the public discussion it is complemented by\footnote{As prominent examples may serve the Snowden, Cambridge Analytica, or lately, Pegasus revelations.}) significantly influences business practices. Regulations, such as the GDPR or the California Consumer Privacy Act (CCPA) \cite{ccpa} provide strong regulatory frameworks which are accompanied by landmark case law decisions (such as \enquote{Schrems II}\footnote{See \url{https://curia.europa.eu/juris/document/document.jsf?text=&docid=228677&doclang=en}}). Eventually, the legal perspective of privacy boils down to several foundational principles (e.g., transparency, data minimization, or accountability) which have been accepted as common ground (inter alia, \cite{cavoukian2009privacy, oecd}). Therefore, in section~\ref{dimensions-cloud-native} we extract the central privacy principles which are encoded in the GDPR to be reflected with the cloud native and agile software development trends laid out above. Before that, we briefly introduce the discipline of privacy engineering.

\subsection{Privacy Engineering}\label{sec:peng} %

Privacy engineering is the discipline of technically addressing the aforementioned privacy principles to protect data subjects and to avoid threats and vulnerabilities (inducing risks) while meeting all functional and non-functional requirements of data controllers and processors. Clearly, this does not only include the operationalization of producing source code, but also encompasses the holistic view on software architecture, business organization and culture including all stakeholders. This perspective led to the umbrella term \textit{Privacy by Design and By Default} \cite{cavoukian2009privacy, gurses2011engineering, spiekermann2012challenges, hansen2016data}. From a legal perspective, privacy engineering is motivated through said motto in Art. 25 GDPR. Controllers, therefore, have to take into account the \enquote{state of the art, the cost of implementation and the nature, scope, context and purposes of processing as well as the risks of varying likelihood and severity} of processing personal data. Further, \enquote{appropriate technical and organisational measures} need to be implemented. As a consequence, there is a steady and momentous incentive for building applicable technical components. Since they may advance the state of the art, they then have to be used by data controllers in practice to protect data subjects. Naturally, when exactly the state of the art might be significantly advanced is questionable from case to case. However, the GDPR, for instance, enables certification procedures in Art. 42, which also take into consideration the differences between dominant economic players and small and medium-sized enterprises. Additionally, among others, the European Data Protection Board, constantly publishes guidelines and recommendations which are clear indicators on compliant technical and organizations measures. Likewise, other civic or research institutions provide their expertise to the public.

Focusing on the implementation, Privacy Enhancing Technologies (PETs) are subject to the core of privacy engineering research. With each generation of new technologies, the conceptual frameworks further matured:
From early visions \cite{goldberg1997privacy},
over elaborated strategies for software architecture in practice \cite{hoepmann2014strategies, heurix2015taxonmy, hansen2016data} and related privacy patterns\footnote{See \url{https://privacypatterns.org/}.},
to topical challenges of software engineering and service architectures \cite{kostova2020privacy}.

Reputed early projects such as Cranor's P3P \cite{cranor2002web} or the European PRIME \cite{hansen2004privacy} and PrimeLife \cite{pfitzmann2011primelife} catalysed the discourse around PETs further. More recent projects then focused on privacy and especially transparency, also in distributed contexts (e.g.,
Privacy \& Us\footnote{See \url{https://privacyus.eu/}},
PRISMACloud\footnote{See \url{https://prismacloud.eu/}},
SPECIAL\footnote{See \url{https://specialprivacy.ercim.eu/}},
or DaSKITA\footnote{See \url{https://daskita.github.io/}}). While many approaches focus on (not less important) data subject facing technologies (such as privacy dashboards), key advances that keep pace with the rising complexity of distributed cloud native systems are hard to identify.  

Still, product managers and software engineers are ill-equipped with the right tools to put privacy by design in practice. Studies show, that there is a fundamental responsibility issue among engineers \cite{spiekermann2019inside}.

\pagebreak

Although the majority of them is aware of the threats of non-compliant software systems and the potential harm they could produce to data subjects, they lack the means to proactively implement countermeasures against attack vectors or the ethical design of IT infrastructures \cite{spiekermann2019inside}. Further, extensive literature review reveals that there are \textit{(i)} a lack of viable tools and practices for the complete software development cycle, and \textit{(ii)} misconceptions when such implementations achieve their goals \cite{alslais2020privacy}.

As introduced above, the way software is developed has fundamentally changed (\enquote{The Agile Turn}). Traditional shrink-wrap products are to be replaced by interconnected online service offerings powered by cloud native architectures \cite{gurses2018agile}. This, in turn, makes it inevitable to rethink both, the complexity of interrelations of data processors and the functional and non-functional requirements the future generation of PETs needs to address. The same is true for the resulting automation potentials, e.g., within data protection impact assessments \cite{zimmermann2020}. Furthermore, cloud native engineering is constantly in flux and will be extended through IoT and fog computing scenarios \cite{pallas2020fog}. Therefore, we continue examining which dimensions cloud native privacy engineering is subject to in the following section.

\section{Dimensions of Cloud Native Privacy Engineering} \label{dimensions-cloud-native} %

In the following section, we propose a cloud native privacy engineering matrix, that illustrates conceptual dimensions, which will be exemplified by subsequent use case scenarios.

First, we reiterate the importance of regulatory frameworks such as the GDPR \cite{gdpr} or the CCPA \cite{ccpa} in the context of privacy-aware cloud systems -- we refer to \matrixElement{matrixYellow}{legislation}. Through further legislative proposals such as the European ePrivacy Regulation\footnote{See \url{https://eur-lex.europa.eu/legal-content/EN/TXT/?uri=CELEX\%3A52017PC0010}.}, Data Governance Act\footnote{See \url{https://eur-lex.europa.eu/legal-content/EN/TXT/?uri=CELEX\%3A52020PC0767}.}, and the Digital Services Act\footnote{See \url{https://eur-lex.europa.eu/legal-content/en/TXT/?uri=COM:2020:825:FIN}.} the future guidelines will be complemented. Together with evolving social norms and expectations or professional privacy threat analysis frameworks, such as LINDDUN \cite{deng2011privacy}, these will and already are highly influencing the compliance strategies of enterprises. Therefore, the discipline of privacy engineering has to keep track of all these legal requirements to be implemented in their software products.
\vskip2pt
Second, enterprises are changing their \matrixElement{matrixYellow}{organization} through more innovative workforce structures. On the one hand, many firms are no longer just supported by software, but software development is at the core of their business activity. With these changes come shifts in personnel and governance structures, roles and responsibilities, and more flexible methods of operation. This is why, from a business perspective, established models to integrate privacy need to be reviewed. These concerns are of utmost importance for decision-makers and strategists within companies to align with the aforementioned regulatory requirements (i.e. in order to avoid penalties), but also to keep being competitive. In the cloud native context, this includes, for instance, make-or-buy or vendor lock-in decisions with regard to (multi- / hybrid-) cloud computing infrastructure or (external) privacy consulting.
\vskip2pt
Third, we emphasize a \matrixElement{matrixYellow}{process}-related dimension. Closely related to the organizational questions are the handling of effective communication and clear privacy by policy \cite{spiekermannEngineering} responsibilities. From a computer science and engineering perspective, technical components are aspired to automate as many things as possible. As we will see later on, the smart implementation of privacy-related tools into the continuous integration and deployment (CI/CD) workflows can greatly heighten the level of data protection. However, \enquote{purely technical approaches might prove insufficient for aligning nuanced legal policies with engineering artifacts} \cite{gurses2016privacy}. As a consequence, engineers need to be engaged and cherished for their individual contributions to all cloud native privacy engineering efforts. This can be done through a supportive and efficient culture, incentive schemes and, most importantly, developer-centric privacy engineering solutions. These are primarily characterized through developer-friendliness (including intuitive usage, appropriate documentation etc.) and low implementation overhead \cite{gruenewald2021, pallas2020pbac}. At the same time, already established cloud native tooling provides tremendous potential to be unlocked for \textit{(i)} aligning with privacy law, \textit{(ii)} supporting organizational efficacy, and \textit{(iii)} automating many steps of the process of dealing with hundreds of services. All of these reflect the highly-specific perspectives driven by the business model and implementation of fulfillment processes of a data controller.

Furthermore, cloud native engineering is heavily focused on the specifics of (at least) three different layers. Usually, these layers are denoted as Infrastructure, Platform, and Software as a Service (XaaS). These terms emphasize the share between self-managed and fully provided solutions by the cloud provider. Since we are discussing software development in general, we rename \enquote{Software} to \enquote{Application} layer to avoid confusion. Thus, all major cloud vendors offer\footnote{Note that some of these example attributions may differ in details depending on their concrete system design. Some of the abstraction levels also increasingly blur together.} three layers:

\newpage

\begin{itemize}\itemsep2pt
    \item \matrixElement{matrixGreen}{Infrastructure} that consists of compute, storage, and network resources (virtual and/or pooled)
    \begin{itemize}
        \item \textit{Examples: Virtual machines, Storage buckets, Software Defined Networks}
    \end{itemize}
    \item \matrixElement{matrixGreen}{Platform} for building, testing, deploying, running, and scaling services on managed infrastructure
    \begin{itemize}
        \item \textit{Examples: Container orchestrators, Serverless / Functions as a Service, Pre-trained machine learning environments, Managed databases, Elastic load balancers}
    \end{itemize}
    \item \matrixElement{matrixGreen}{Application} that is handling the business logic and may contain several user or application programming interfaces.
    \begin{itemize}
        \item \textit{Examples: Depending on the business scenario, any application written in any programming language incl. interface and communication specifications.}
    \end{itemize}
\end{itemize}

All of the latter are building blocks for large-scale data processing. From this follows, privacy engineering needs a bouquet of solutions to cope with the different deployment models and configurations of cloud native architecture, since personal data is processed in many different ways. We have now identified the first six dimensions of cloud native privacy engineering. Three of them (Legislation, Organization, and Process) are addressing mainly the external factors privacy engineers are influenced by.

Oriented orthogonally to the dimensions already mentioned, we will therefore now add 10 more to complete the proposed view of cloud native privacy engineering. All of the following ones are distilled from both literature and the GDPR, who we denote as essential privacy principles. Note that none of the following principles is new per se, however, it is of utmost importance to see them in conjunction with the aforementioned cloud engineering layers of abstraction. For an in-depth study, we refer to extensive related work \cite{cavoukian2009privacy, voigt2017eu, kunerbackground}. We only list them very briefly for the sake of simplicity:

\begin{itemize}\itemsep2pt
    \item \matrixElement{matrixBlue}{Lawfulness} (Art. 5(1a), 6--11 GDPR) comprises the prohibition of all personal data processing activities \textit{unless} there is one of the well-defined permission options present (e.g. consent).

    \item \matrixElement{matrixBlue}{Fairness} (Art. 5(1a) GDPR) refers to proportionality between interests and necessities of both data controllers and data subjects. Moreover, it can be interpreted as procedural fairness which includes timeliness or burden of care \cite{clifford2018fairness}. Fairness is also an umbrella term for multiple concepts as defined by the OCED guidelines \cite{oecd} and the Fair Information Practices \cite{ftc2000}.\footnote{Note, although fairness \enquote{remains under-defined from a legal perspective}, it still has to be considered in explicit design trade-offs; see also \cite{finck2021}.}
    
    \item \matrixElement{matrixBlue}{Transparency} (Art. 5(1a), 12, 13, 14, 30 GDPR) includes transparent information, communication and modalities for the exercise of data subjects and the respective obligations for data controllers or processors which allows independent verification and enables trust \cite{cavoukian2009privacy}.    

    \item \matrixElement{matrixBlue}{Accountability} (Art. 5(2), 24 GDPR) entails the responsibility and ability for demonstration of compliance with all the other principles. Therefore, it is closely related to enforcement and audit strategies of supervisory authorities.   
    
    \item \matrixElement{matrixBlue}{Purpose limitation} (Art. 5(1b) GDPR) requires specific, explicit, and legitimate purpose specifications. This prohibits overly broad statements and data processing upon retrospective amendments or further incompatible processing with the initially stated purpose. %

    \item \matrixElement{matrixBlue}{Data minimization} (Art. 5(1c) GDPR) limits the collection of personal data for further processing. Frequent tactics are excluding, selecting, stripping, perturbating, and deleting personal data as much as possible \cite{hoepmann2014strategies}. Possible safeguards include anonymization and (to a limited degree) pseudonymization.

    \item \matrixElement{matrixBlue}{Accuracy} (Art. 5(1d) GDPR) determines that all personal data are to be kept up-to-date and correct. Therefore, data subjects have the right to rectification (Art. 16), which is important to reduce possible algorithmic discrimination because of false assumptions.

    \item \matrixElement{matrixBlue}{Storage limitation} (Art. 5(1e) GDPR) specifies period for which personal data can be processed. This period is strongly coupled to the lawfulness and the specific purpose for which the processing is permitted.
    
    \item \matrixElement{matrixBlue}{Security} (Art. 5(1f), 32 GDPR) safeguards against unauthorized and unlawful data processing. The technical and organizational measures need to ensure confidentiality, integrity, and availability (CIA triad) \cite{agarwal2011security}. 

    \item \matrixElement{matrixBlue}{Access \& Data portability} (Art. 15, 20 GDPR) refer to all data subjects' right to get a copy of all personal data relating to them. Closely related, the GDPR guarantees the freedom - where technically feasible - to transmit their personal data from one controller to another. The latter also enables a (in theory) effective mean against dominant market positions \cite{diker2017right}.
\end{itemize}

\begin{figure}[H]
    \centering
    \includegraphics[width=0.90\linewidth]{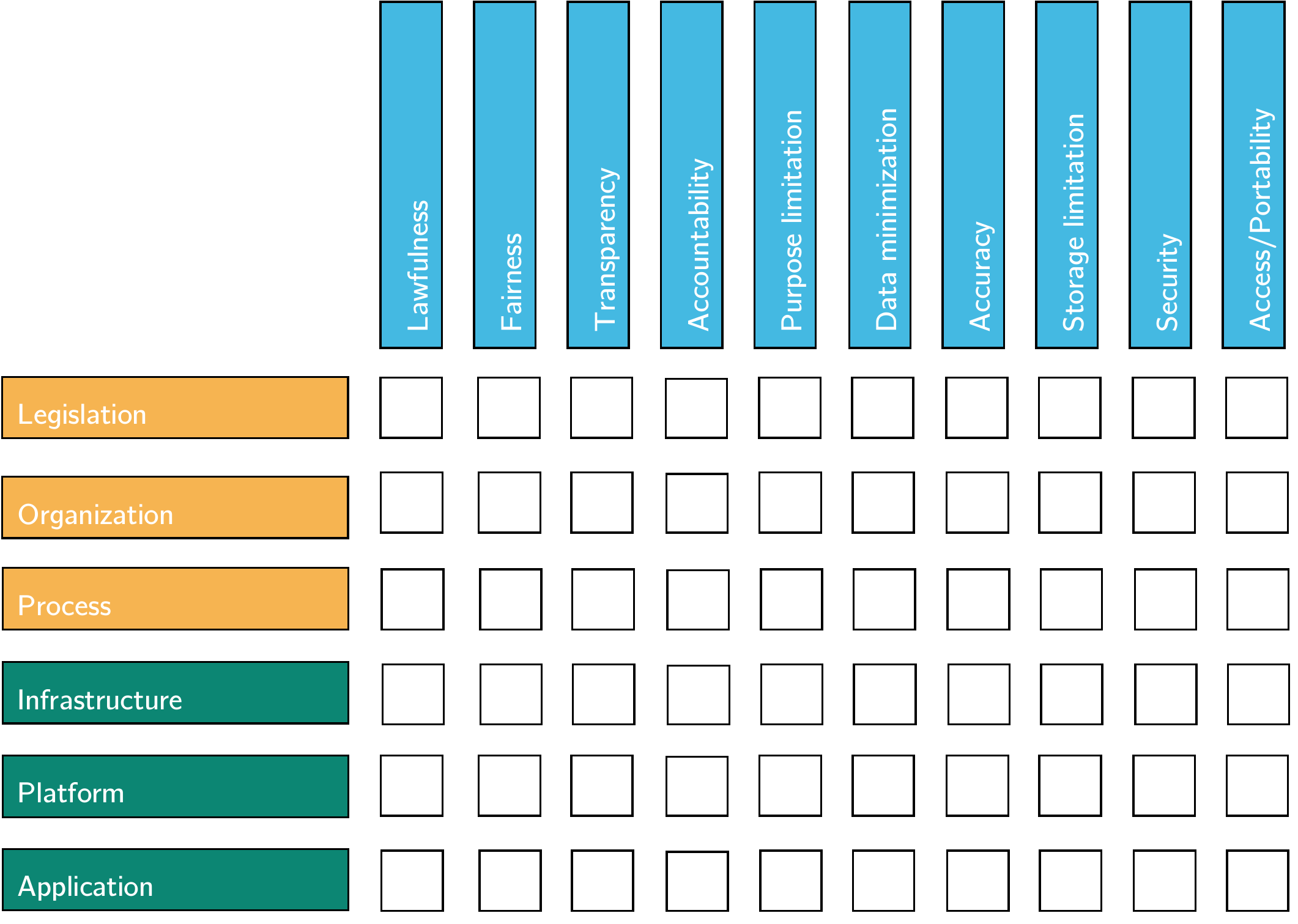}
    \caption{Dimensions of Cloud Native Privacy Engineering}
    \label{fig:lopi}
\end{figure}

Privacy by design needs to target a positive-sum, not zero-sum to unfold its real societal impact \cite{cavoukian2009privacy}. Although within systems engineering trade-offs need to be discussed during the development process, the ultimate goal has to be to align with \textit{all} the privacy principles best. In this context, we also acknowledge the classifications of privacy engineering \textit{by architecture} \cite{spiekermannEngineering}, \textit{policy} \cite{spiekermannEngineering}, and \textit{interaction} \cite{gurses2016privacy} which clear the mist for evaluating proposed systems. In addition, we can contextualize (again) the privacy design strategies \cite{hoepmann2014strategies} that can be directly mapped to many of the resulting matrix elements which are depicted in Fig.~\ref{fig:lopi}.

We now examine two different use cases, illustrating the range of different privacy engineering mechanisms:\newline

\begin{wrapfigure}{T}{0.33\textwidth}
    \includegraphics[width=\linewidth]{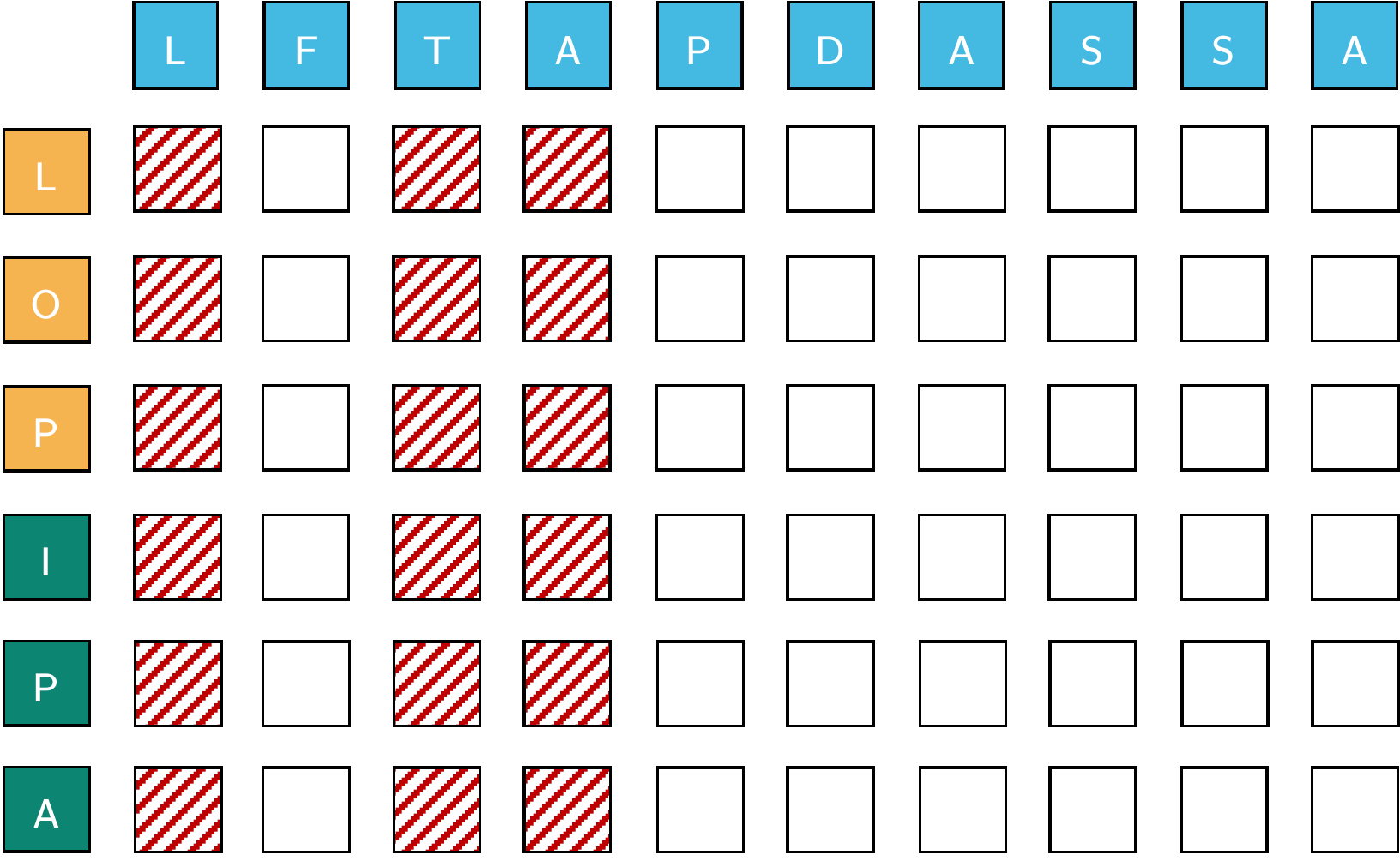}
\end{wrapfigure}
\noindent \textbf{Use case 1.} Transparency in large service-based cloud architectures is key to strengthen data subjects' level of informedness. Traditionally, written privacy policies try to convey transparency information in legalese language. However, they are not only hard to understand for users, but also incompatible with agile development practices, as they -- by design -- cannot be changed multiple times per day. Cloud native architectures, in turn, need a machine-readable representation and additional tooling for processing said transparency information in order to describe the multitude of services in real-time. TILT \cite{gruenewald2021} and TIRA \cite{gruenewald2021tira}, as technical mechanisms, address this issue being explicitly tailored to large-scale cloud native systems, agile development practices, and the legal requirements. Consequently, the proposed policy language and programming toolkit of TILT, and the OpenAPI extension and dashboard of TIRA address transparency, accountability, and lawfulness on many different levels.\newline

 \newpage

\begin{wrapfigure}{T}{0.33\textwidth}
    \includegraphics[width=\linewidth]{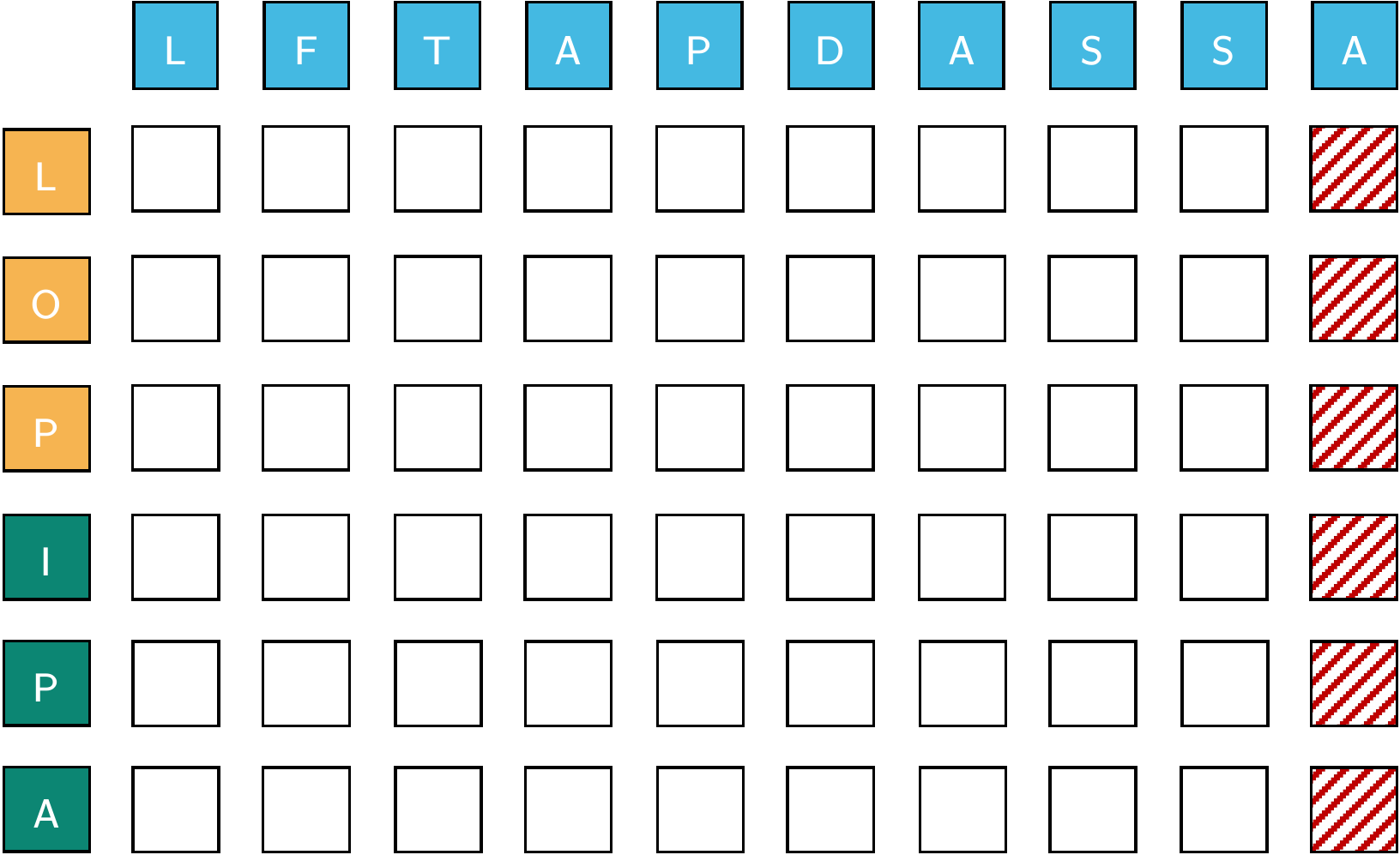}
\end{wrapfigure}
\noindent \textbf{Use case 2.} The Right to Data Portability (RtDP) is still uncharted territory in real-world systems. At least, many data controllers provide so-called takeouts\footnote{E.g. Google's takeout under \url{https://takeout.google.com/settings/takeout}.} for semi-automatically fulfilling the right to access according to Art. 15 GDPR. Also the CCPA clearly specifies in Sec.~III that data controllers \enquote{shall promptly take steps to disclose and deliver, free of charge to the consumer, the personal information required}. Closely related, Art. 20 GDPR states the right to data portability. As a consequence, the data also has to be provided in a machine-readable format. However, the automatic transfer of all personal data from controller A to B is still somewhat disregarded. At least, one major PET has been proposed by a consortium of big technology companies, namely the Data Transfer Project (DTP)\footnote{See \url{https://datatransferproject.dev/}.}. The DTP addresses the RtDP through three main components. First, there are several data models that can be extended by the community, and are to be used for describing the personal data to be transferred. Next, they propose company-specific adapters for authentication and how to communicate with the provider's core infrastructure (preferably through well-defined APIs). Third, they connect these components through various middleware components enabling in-transit encryption or failure handling. The project is in an experimental state, however, it is a serious attempt to enable the RtDP. Notably, the tool is built using common cloud native techniques such as containerization and well-defined web APIs for developer-friendly integration at application level.\newline %

As we can infer from these two examples, cloud native privacy engineering is still a difficult endeavour. On the one hand, there is no such option as free lunch, since not a single or two tools can possibly cover the complete range of the dimensions at hand. Secondly, a remaining question is as to whether a PET is considered as \enquote{appropriate} measure. Calculating the risk of a password brute-force attack is fairly easy, while, in comparison, measuring an adequate level of \textit{fair} or \textit{transparent} data processing is an unsolved problem. Notwithstanding, by the help of the proposed model, we can now compare different architectures by checking how sparse or dense the matrix is filled. As a rule of thumb, the more privacy principles at different levels are met (indicated by a colored matrix element), the better is the overall rating. Ideal privacy engineering solutions then cover complete columns or even span rows. In contrast, a system described by a sparse matrix faces a substantial need for remedial action. In a second step, case-specific data protection impact assessments (DPIAs) should be carried out. By its very nature, the level of ensured privacy cannot be put into a single evaluation model. However, evidence-based experiments and research shall complement the discussion: On the one hand, we need to consider the cost of implementation efforts according to Art. 25(1) in relation to the processing at hand. On the other hand, we argue that all phases of development and operations need to be taken into account. Consequently, we need empirical studies for various kinds of PETs relating to all dimensions of cloud native privacy engineering. Having these, we can better compare and evaluate complete systems w.r.t. to architecture, engineering, and management.

After having discussed the dimensions of cloud native privacy engineering, we head over towards the software development cycle to demonstrate the implementation in practice.

\section{DevPrivOps: Privacy Engineering in Practice} \label{dev-priv-ops} %

In this section, we suggest an enhanced \textit{DevPrivOps} lifecycle complementing the model of \cite{sion2020neverending}, that illustrates how privacy can be ensured in cloud native architectures and through which tools the privacy-friendly and agile development of large-scale service infrastructures can be exemplified.

\begin{figure}[h]
    \centering
    \includegraphics[width=0.85\linewidth]{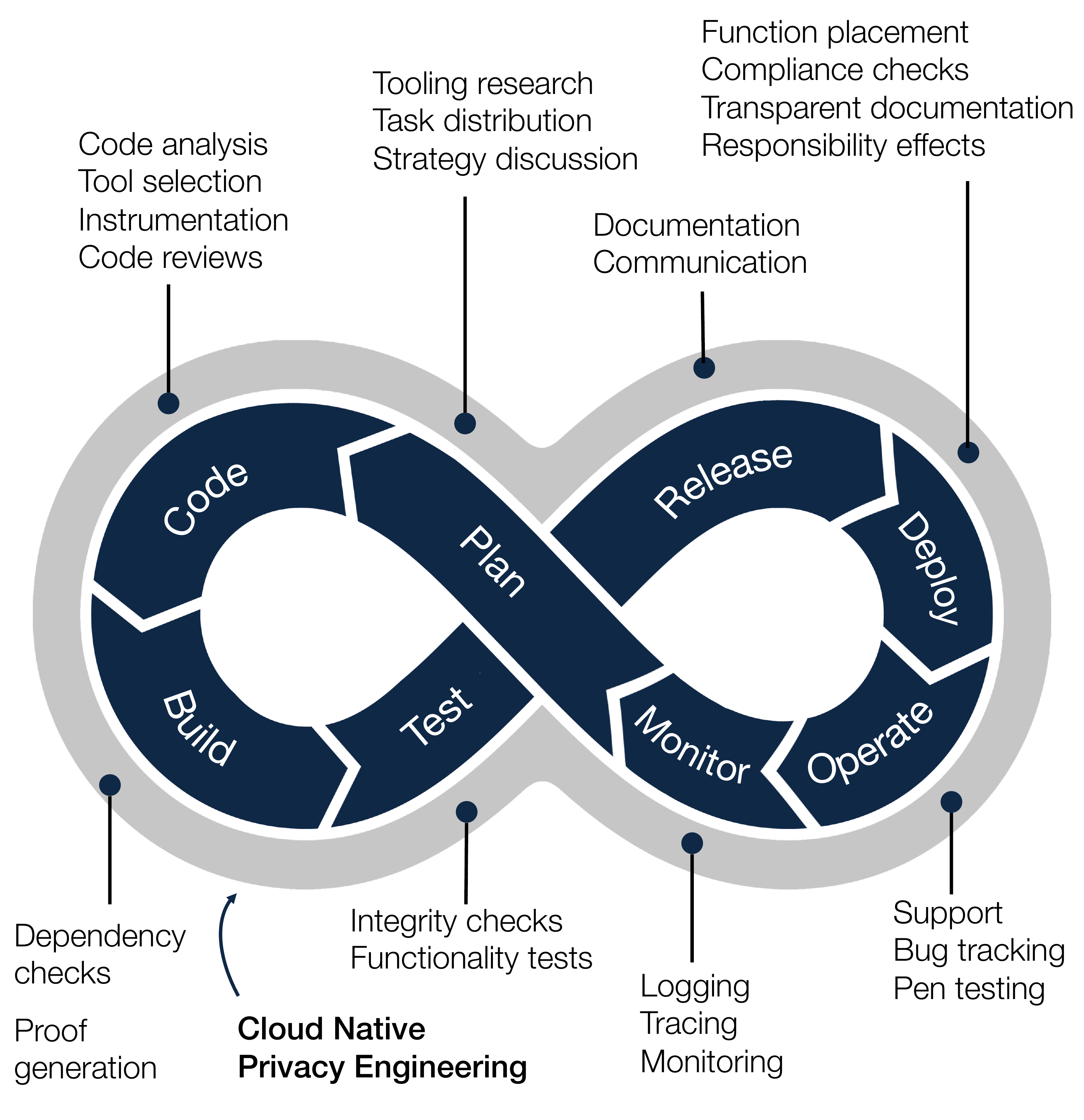}
    \caption{Continuous DevPrivOps software development lifecycle.}
    \label{fig:DevPrivOps}
\end{figure}

DevOps emphasizes cross-functional collaboration to operate systems and accelerate delivery of any occurring changes \cite{dyck2015definition}. For this purpose, it is practiced as a software development culture that integrates the following eight phases conducted in an endless cycle \cite{yarlagadda2021devops}. We will explain them briefly in our own words (for long-reads we recommend \cite{bass2015devops, stahl2017continuous}). Additionally, we will hint at tangible activities that complement the phase with cloud native privacy engineering tactics. Therefore, we can now introduce a DevPrivOps lifecycle, that consists of the already established DevOps loop (depicted in blue) and an enveloping \enquote{ring} that illustrates the possibility to add privacy-related activities in every phase (cf. Fig.~\ref{fig:DevPrivOps}).

First, the lifecycle is initialized by a \textbf{planning} phase. Working with agile project management tools, this could be a scrum planning phase, in which the tasks for the next sprint are to be defined. Moreover, this phase serves as a checkpoint to plan either a new functionality or fixes and enhancements to an existing one. Changes can, e.g., be prioritized based on the developer's skill or the strategic business reason why a change is requested. In traditional software engineering, the plan phase is comparable to the requirements engineering phase, in which all functional and non-functional items are to be collected. Within the planning phase, it is convenient for the team to discuss which privacy pattern or design strategy (see section~\ref{sec:peng}) to employ. This phase may also entail the threat modelling or risk analysis to decide which technologies fit best.

Second, writing of source \textbf{code} begins. This activity is not meant be limited to programming in the general purpose languages at hand, but can also be used to write configuration files, infrastructure as code definitions \cite{artac2017devops}, test cases, API specifications or database queries (non-exhaustive list). Coding is assisted by integrated development environments (IDEs), a collection of tools to assist writing code, debugging, reading documentation and so on. With regard to privacy engineering, this phase is used to employ libraries or plugging in components that feature a design goal. For the security dimension one would, e.g., choose the encryption cipher suite and library. When focusing on transparency, all personal data indicators \cite{gruenewald2021tira} would be documented (which also streamlines auditability) or (manual or automatic) instrumentation for logging, tracing, and monitoring tools would be added. Basically, this phase is crucial for every processing activity. Some IDEs automatically hint at uncatched exceptions, possible SQL injections, non-documented function parameters, missing type checking and many possible other security flaws in the source code \cite{li2019evaluation}. In addition, version control systems are used to organize multiple developers working on the same files in different development branches. These can further be used to review code changes by another team member. This enables shared responsibilities and better code quality. 

Third, the application is built using \textbf{build} automation tools. These tools help to check if all external and internal dependencies can be resolved or supervise the compilation process of respective programming languages. With regard to security, outdated versions of external libraries could be identified. Taking the data minimization and purpose limitation dimensions as examples, the tools can assist in building different versions for disparate target groups. For instance, if the business model contains a paid version without targeted advertising, the build automation could exclude third party tracking functionalities. Besides, in trustless setups, for instance, zero-knowledge proofs are generated in order to keep sensitive information private \cite{eberhardt2018zokrates}.  

After the build phase, automatic \textbf{tests} are executed. Software testing can be an exhaustive task that includes thousands of test cases. Using the testing phase for privacy-related tasks, the test suite can check several functions with different inputs if the expected accuracy or integrity are ensured. Test data sets can be used to check different behaviours or the correct calculation of obfuscation mechanisms. For instance, parts of a threat analysis and management can be automated in a CI pipeline \cite{sion2021automated}. Generally, integration tests can also be used to check the platform or infrastructure in various dimensions (esp. with security in mind) \cite{hsu2018hands}.

Next, the changes are \textbf{release}d. Therefore, DevOps engineers automate an integration pipeline that is executed automatically. Such a pipeline may again carry out tests on different target platforms and then create a package for later delivery. Privacy engineering can play a role here again, e.g., by executing integrity checks or adding transparency-related information that can be generated out of automated analysis tasks. With this phase we leave the rather development-focused phases and enter the operation part of the loop.

Afterwards, the software is \textbf{deploy}ed to compute, storage, and network resources, provided by the cloud infrastructure. In automated scenarios, this may include the decision which (virtual) machine is used or at which edge device a container is placed (in a fog computing scenario). In highly distributed scenario these decisions can have a huge impact on the regulatory obligations that apply. Thinking about services that are deployed to a data center in a third country; this has direct implications on transparency, accountability or security dimensions \cite{crabtree2018building}. However, also organizational responsibilities naturally change when software is deployed to infrastructure at different locations.

Subsequently, \textbf{operating} the software is a key task for the responsible team. This does not only include to keep running the software technically, but may also include process-related activities like internal support or bug tracking. From a privacy perspective, security-related tasks such as pen testing or research for vulnerabilities are important. Moreover, during operation potentially lots of personal data is accessed, changed, added, or deleted. These activities need to be observed and cross-checked with the prior-made assumptions about, e.g., to which degree $k$/$k_s$-anonymity \cite{sweeney2002k, redcastle2021}, $\ell$-diversity \cite{machanavajjhala2007diversity} or $\varepsilon$-differential \cite{lee2011much} privacy can be guaranteed in real-world scenarios. In another dimension, data accuracy could be validated after each change.

Thereafter, the \textbf{monitoring} phase is entered. During this phase, cloud native architectures are watched using observability techniques. The most common tools perform logging, distributed tracing and collecting metrics \cite{picoreti2018multilevel}. These tools can also be used to achieve a higher level of privacy. First, logging helps to build an accountable system, since the controller can historically demonstrate that the system worked as intended by keeping the records of processing activity \cite{felici2013accountability}. Secondly, distributed tracing can be used to observe service compositions. Thus, a data controller has full transparency over all personal data processing and can provide a summary to the data subject or supervisory authority in real-time. Moreover, all joint controllerships or de-facto processors are automatically detected independently from what was manually documented before. Additionally, purpose limitation can be guaranteed when there is a \enquote{watchdog} that detects unwanted or unlawful behaviour. Third, by the help of collected metrics we can detect adverse intrusions and therefore threats for personal data leakage. At the same time, key performance indicators allow to prove that data access or portability tasks were timely executed. As shown, this phase is exceptionally well suitable for all different kinds of cloud native privacy engineering. 

In this section we showed how privacy engineering can be integrated into all phases of DevOps engineering. The term \textit{DevPrivOps} was mentioned first in a recent position paper \cite{sion2020neverending}. With this paper we want to further coin the term with regard to cloud computing environments and the chances of cloud-native tools to implement all privacy principles in practice. So far, the term DevOps was only augmented as \textit{DevSecOps}. However, this perspective does not reflect the complexity of the privacy engineering discipline as a whole.

\section{Discussion and Conclusion} \label{discussion} %

Future privacy engineering tools need to be in line with the actual givens of practical information systems engineering. Without doubt, conceptual models provide us with guidelines for a better architecture of real-world systems. At the same time, a reality check is necessary to align with the environment of industry-grade cloud native computing. So far, we observed that the goal of privacy by design can then be reached when the software development lifecycle is focused on all dimensions as laid out in section~\ref{dimensions-cloud-native}. 

In this paper, malicious practices from within the corporation were not considered. It is evident that appropriate measures must also be taken in this regard. These are, however, primarily of an organizational nature. Consequently, clear processes and, above all, automated tests can also be effective protective safeguards. Manual manipulation away from the log is made significantly more difficult by DevPrivOps practices, e.g., when CI/CD pipelines alert or stop non-compliant code from running in production.

To the best of our knowledge, there have been no studies on the acceptance of privacy engineering methods that encompass all the dimensions identified above. Rather, there is the impression that priority is given to security- and data minimization related measures, while most other principles are neglected. To counteract this, further incentive models and easy-to-integrate technical options need to be developed. In this paper, we have provided first suggestions; in turn, a more comprehensive list of options for implementing privacy by design needs to be extracted from existing DevOps implementations.

In the same vein, we need evaluation methods for each of the elements in the cloud native privacy engineering matrix. For the development of these, we can first borrow ideas from both legal and technical methodology. Then, we need to carry out a cross-disciplinary discourse on the exact design of said approaches. This process is considered future work needed to be coming up next. Having these evaluation methods set up, we then can also better evaluate the \enquote{level of coverage} within each matrix element (possibly indicated by a filling level instead of hatching).%

Finally, technological possibilities continue to grow at a rapid pace. With increasing connectivity through powerful mobile networks, the number of internet-enabled devices that will process personal data is exploding. For computing approaches such as fog computing and (I)IoT (Industrial Internet of Things), appropriate strategies need to be developed and tested to effectively implement the above privacy dimensions. However, they can also be beneficial and helpful for said principles \cite{pallas2020fog}.

So far, this work has presented two key models that bring privacy engineering by design and by default closer to the realities of information systems engineering. At the same time, it is intended to improve these designs in future iterations -- just as prompted by the infinite loop presented above.

\section*{Acknowledgements}
\begin{footnotesize}
The work behind this paper was partially conducted within the project DaSKITA (Data sovereignty through AI-based transparency and access), supported under grant no. 28V2307A19 by funds of the Federal  Ministry of Justice and Consumer Protection (BMJV) based on a decision of the Parliament of the Federal Republic of Germany via the Federal Office for Agriculture and Food (BLE) under the innovation support program.
\end{footnotesize}

\bibliographystyle{splncs04}
\bibliography{bibliography.bib}

\begin{thebibliography}{10}
\providecommand{\url}[1]{\texttt{#1}}
\providecommand{\urlprefix}{URL }
\providecommand{\doi}[1]{https://doi.org/#1}

\bibitem{agarwal2011security}
Agarwal, A., Agarwal, A.: The security risks associated with cloud computing.
  International Journal of Computer Applications in Engineering Sciences
  \textbf{1},  257--259 (2011)

\bibitem{alslais2020privacy}
Al-Slais, Y.: Privacy engineering methodologies: A survey. In: 2020
  International Conference on Innovation and Intelligence for Informatics,
  Computing and Technologies (3ICT). pp.~1--6 (2020).
  \doi{10.1109/3ICT51146.2020.9311949}

\bibitem{artac2017devops}
Artac, M., Borovssak, T., Di~Nitto, E., Guerriero, M., Tamburri, D.A.: Devops:
  {Introducing} infrastructure-as-code. In: 2017 IEEE/ACM 39th International
  Conference on Software Engineering Companion (ICSE-C). pp. 497--498. IEEE
  (2017)

\bibitem{balalaie2016microservices}
Balalaie, A., Heydarnoori, A., Jamshidi, P.: Microservices architecture enables
  devops: Migration to a cloud-native architecture. IEEE Software
  \textbf{33}(3),  42--52 (2016)

\bibitem{bass2015devops}
Bass, L., Weber, I., Zhu, L.: {{DevOps}}: {{A Software Architect}}'s
  {{Perspective}}. {Addison-Wesley} (2015)

\bibitem{bednar}
Bednar, K., Spiekermann, S., Langheinrich, M.: {Engineering Privacy by Design:
  Are engineers ready to live up to the challenge?} The Information Society
  \textbf{35}(3),  122--142 (2019). \doi{10.1080/01972243.2019.1583296}

\bibitem{cavoukian2020}
Cavoukian, A.: Understanding how to implement privacy by design, one step at a
  time. IEEE Consumer Electronics Magazine  \textbf{9}(2),  78--82 (2020).
  \doi{10.1109/MCE.2019.2953739}

\bibitem{cavoukian2009privacy}
Cavoukian, A., et~al.: {Privacy by design: The 7 foundational principles}.
  Information and privacy commissioner of Ontario, Canada  \textbf{5}, ~12
  (2009)

\bibitem{clifford2018fairness}
Clifford, D., Ausloos, J.: {Data Protection and the Role of Fairness}. Yearbook
  of European Law  \textbf{37},  130--187 (08 2018). \doi{10.1093/yel/yey004},
  \url{https://doi.org/10.1093/yel/yey004}

\bibitem{cncf}
{Cloud Native Computing Foundation (CNCF)}: {Cloud Native Definition v1.0}
  (2018), \url{https://github.com/cncf/toc/blob/main/DEFINITION.md}

\bibitem{ccpa}
Code, C.C.: California consumer privacy act ({CCPA}) (2018)

\bibitem{crabtree2018building}
Crabtree, A., Lodge, T., Colley, J., Greenhalgh, C., Glover, K., Haddadi, H.,
  Amar, Y., Mortier, R., Li, Q., Moore, J., et~al.: Building accountability
  into the {Internet of Things}: {The IoT Databox} model. Journal of Reliable
  Intelligent Environments  \textbf{4}(1),  39--55 (2018)

\bibitem{cranor2002web}
Cranor, L.F.: Web privacy with {P3P}. O'Reilly Media, Inc. (2002)

\bibitem{deng2011privacy}
Deng, M., Wuyts, K., Scandariato, R., Preneel, B., Joosen, W.: A privacy threat
  analysis framework: supporting the elicitation and fulfillment of privacy
  requirements. Requirements Engineering  \textbf{16}(1),  3--32 (2011)

\bibitem{diker2017right}
Diker~Vanberg, A., {\"U}nver, M.B.: The right to data portability in the gdpr
  and eu competition law: odd couple or dynamic duo? European Journal of Law
  and Technology  \textbf{8}(1) (2017)

\bibitem{dragoni2017microservices}
Dragoni, N., Giallorenzo, S., Lafuente, A.L., Mazzara, M., Montesi, F.,
  Mustafin, R., Safina, L.: Microservices: yesterday, today, and tomorrow.
  Present and ulterior software engineering pp. 195--216 (2017)

\bibitem{dyck2015definition}
Dyck, A., Penners, R., Lichter, H.: Towards definitions for release engineering
  and devops. In: 2015 IEEE/ACM 3rd International Workshop on Release
  Engineering. pp.~3--3 (2015). \doi{10.1109/RELENG.2015.10}

\bibitem{eberhardt2018zokrates}
Eberhardt, J., Tai, S.: {ZoKrates} - {Scalable Privacy-Preserving Off-Chain
  Computations}. In: IEEE International Conference on Blockchain. pp.
  1084--1091. IEEE (2018)

\bibitem{erich2017qualitative}
Erich, F., Amrit, C., Daneva, M.: A qualitative study of {DevOps} usage in
  practice. Journal of Software: Evolution and Process  \textbf{29}(6),  e1885
  (2017)

\bibitem{gdpr}
{{European Parliament and Council of the European Union}}: {Regulation ({EU})
  2016/679 of 27 {April} 2016. {General Data Protection Regulation}} (2018)

\bibitem{ftc2000}
{Federal Trade Commission}: Privacy online: Fair information practices in the
  electronic marketplace (2000),
  \url{https://www.ftc.gov/reports/privacy-online-fair-information-
  practices-electronic-marketplace-federal-trade-commission}

\bibitem{felici2013accountability}
Felici, M., Koulouris, T., Pearson, S.: Accountability for data governance in
  cloud ecosystems. In: 2013 IEEE 5th International Conference on Cloud
  Computing Technology and Science. vol.~2, pp. 327--332. IEEE (2013)

\bibitem{finck2021}
Finck, M., Biega, A.J.: Reviving purpose limitation and data minimisation in
  data-driven systems. Technology and Regulation  \textbf{2021},  44--61 (Aug
  2021), \url{https://techreg.org/index.php/techreg/article/view/63}

\bibitem{gannon2017cloud}
Gannon, D., Barga, R., Sundaresan, N.: Cloud-native applications. IEEE Cloud
  Computing  \textbf{4}(5),  16--21 (2017)

\bibitem{gill2019transformative}
Gill, S.S., Tuli, S., Xu, M., Singh, I., Singh, K.V., Lindsay, D., Tuli, S.,
  Smirnova, D., Singh, M., Jain, U., et~al.: Transformative effects of iot,
  blockchain and artificial intelligence on cloud computing: Evolution, vision,
  trends and open challenges. Internet of Things  \textbf{8},  100118 (2019)

\bibitem{goldberg1997privacy}
Goldberg, I., Wagner, D., Brewer, E.: Privacy-enhancing technologies for the
  internet. In: Proceedings IEEE COMPCON 97. Digest of Papers. pp. 103--109.
  IEEE (1997)

\bibitem{gruenewald2021}
Grünewald, E., Pallas, F.: {TILT}: A {GDPR-Aligned Transparency Information
  Language and Toolkit for Practical Privacy Engineering}. In: Proceedings of
  the 2021 Conference on Fairness, Accountability, and Transparency. ACM, New
  York, NY, USA (2021). \doi{10.1145/3442188.3445925}

\bibitem{gruenewald2021tira}
Grünewald, E., Wille, P., Pallas, F., Borges, M.C., Ulbricht, M.R.: ~~{TIRA}:
  An {OpenAPI} {Extension} and {Toolbox} for {GDPR} {Transparency} in {RESTful}
  {Architectures}. In: 2021 International Workshop on Privacy Engineering
  (IWPE). IEEE Computer Society (2021)

\bibitem{gurses2016privacy}
G{\"u}rses, S., Del~Alamo, J.M.: Privacy engineering: Shaping an emerging field
  of research and practice. IEEE Security \& Privacy  \textbf{14}(2),  40--46
  (2016)

\bibitem{gurses2011engineering}
G{\"u}rses, S., Troncoso, C., Diaz, C.: Engineering privacy by design.
  Computers, Privacy \& Data Protection  \textbf{14}(3), ~25 (2011)

\bibitem{gurses2018agile}
Gürses, S., van Hoboken, J.: Privacy after the Agile Turn, p. 579–601.
  Cambridge Law Handbooks, Cambridge University Press (2018).
  \doi{10.1017/9781316831960.032}

\bibitem{hansen2016data}
Hansen, M.: Data protection by design and by default {\`a} la {European General
  Data Protection Regulation}. In: IFIP International Summer School on Privacy
  and Identity Management. pp. 27--38. Springer (2016)

\bibitem{hansen2004privacy}
Hansen, M., Berlich, P., Camenisch, J., Clau{\ss}, S., Pfitzmann, A., Waidner,
  M.: Privacy-enhancing identity management. Information security technical
  report  \textbf{9}(1),  35--44 (2004)

\bibitem{heurix2015taxonmy}
Heurix, J., Zimmermann, P., Neubauer, T., Fenz, S.: A taxonomy for privacy
  enhancing technologies. Computers \& Security  \textbf{53},  1--17 (2015).
  \doi{https://doi.org/10.1016/j.cose.2015.05.002}

\bibitem{hoepmann2014strategies}
Hoepman, J.H.: Privacy design strategies. In: Cuppens-Boulahia, N., Cuppens,
  F., Jajodia, S., Abou El~Kalam, A., Sans, T. (eds.) ICT Systems Security and
  Privacy Protection. pp. 446--459. Springer Berlin Heidelberg, Berlin,
  Heidelberg (2014)

\bibitem{hsu2018hands}
Hsu, T.H.C.: Hands-On Security in {DevOps}: Ensure continuous security,
  deployment, and delivery with {DevSecOps}. Packt Publishing Ltd (2018)

\bibitem{kostova2020privacy}
Kostova, B., G{\"u}rses, S., Troncoso, C.: Privacy engineering meets software
  engineering. on the challenges of engineering privacy bydesign. arXiv
  preprint arXiv:2007.08613  (2020)

\bibitem{kratzke2017understanding}
Kratzke, N., Quint, P.C.: Understanding cloud-native applications after 10
  years of cloud computing-a systematic mapping study. Journal of Systems and
  Software  \textbf{126},  1--16 (2017)

\bibitem{kunerbackground}
Kuner, C., Bygrave, L.A., Docksey, C.: Background and evolution of the eu
  general data protection regulation (gdpr). In: The EU General Data Protection
  Regulation (GDPR). Oxford University Press

\bibitem{lee2011much}
Lee, J., Clifton, C.: How much is enough? {Choosing} $\varepsilon$ for
  differential privacy. In: International Conference on Information Security.
  pp. 325--340. Springer (2011)

\bibitem{lenk2009s}
Lenk, A., Klems, M., Nimis, J., Tai, S., Sandholm, T.: What's inside the cloud?
  an architectural map of the cloud landscape. In: 2009 ICSE workshop on
  software engineering challenges of cloud computing. pp. 23--31. IEEE (2009)

\bibitem{li2019evaluation}
Li, J., Beba, S., Karlsen, M.M.: Evaluation of open-source ide plugins for
  detecting security vulnerabilities. In: Proceedings of the Evaluation and
  Assessment on Software Engineering, pp. 200--209 (2019)

\bibitem{machanavajjhala2007diversity}
Machanavajjhala, A., Kifer, D., Gehrke, J., Venkitasubramaniam, M.:
  l-diversity: Privacy beyond k-anonymity. ACM Transactions on Knowledge
  Discovery from Data (TKDD)  \textbf{1}(1), ~3 (2007)

\bibitem{marston2011cloud}
Marston, S., Li, Z., Bandyopadhyay, S., Zhang, J., Ghalsasi, A.: Cloud
  computing—the business perspective. Decision support systems
  \textbf{51}(1),  176--189 (2011)

\bibitem{mell2011nist}
Mell, P., Grance, T., et~al.: The nist definition of cloud computing  (2011)

\bibitem{mulligan2016privacy}
Mulligan, D.K., Koopman, C., Doty, N.: Privacy is an essentially contested
  concept: a multi-dimensional analytic for mapping privacy. Phil. Trans. R.
  Soc. A.  \textbf{374}(2083) (2016).
  \doi{http://dx.doi.org/10.1098/rsta.2016.0118}

\bibitem{nieuwenhuis2018shift}
Nieuwenhuis, L.J., Ehrenhard, M.L., Prause, L.: The shift to cloud computing:
  The impact of disruptive technology on the enterprise software business
  ecosystem. Technological forecasting and social change  \textbf{129},
  308--313 (2018)

\bibitem{oecd}
{OECD}: {{OECD Guidelines}} on the {{Protection}} of {{Privacy}} and
  {{Transborder Flows}} of {{Personal Data}} (1980)

\bibitem{redcastle2021}
Pallas, F., Legler, J., Amslgruber, N., Grünewald, E.: {RedCASTLE}:
  Practically applicable $k_s$-anonymity for {IoT} streaming data at the edge
  in {Node-RED}. In: Proceedings of the 8th International Workshop on
  Middleware and Applications for the Internet of Things. Association for
  Computing Machinery, New York, NY, USA (2021). \doi{10.1145/3493369.3493601}

\bibitem{pallas2020fog}
Pallas, F., Raschke, P., Bermbach, D.: Fog computing as privacy enabler. IEEE
  Internet Computing  \textbf{24}(4),  15--21 (2020).
  \doi{10.1109/MIC.2020.2979161}

\bibitem{pallas2020pbac}
Pallas, F., Ulbricht, M.R., Tai, S., Peikert, T., Reppenhagen, M., Wenzel, D.,
  Wille, P., Wolf, K.: Towards application-layer purpose-based access control.
  In: Proceedings of the 35th Annual ACM Symposium on Applied Computing. pp.
  1288--1296 (2020)

\bibitem{pfitzmann2011primelife}
Pfitzmann, A., Borcea-Pfitzmann, K., Camenisch, J.: Primelife. In: Privacy and
  Identity Management for Life, pp. 5--26. Springer (2011)

\bibitem{picoreti2018multilevel}
Picoreti, R., do~Carmo, A.P., de~Queiroz, F.M., Garcia, A.S., Vassallo, R.F.,
  Simeonidou, D.: Multilevel observability in cloud orchestration. In: 2018
  IEEE 16th Intl Conf on Dependable, Autonomic and Secure Computing, 16th Intl
  Conf on Pervasive Intelligence and Computing, 4th Intl Conf on Big Data
  Intelligence and Computing and Cyber Science and Technology Congress
  (DASC/PiCom/DataCom/CyberSciTech). pp. 776--784. IEEE (2018)

\bibitem{rajkumar2016devops}
Rajkumar, M., Pole, A.K., Adige, V.S., Mahanta, P.: Devops culture and its
  impact on cloud delivery and software development. In: 2016 International
  Conference on Advances in computing, communication, \& automation
  (ICACCA)(Spring). pp.~1--6. IEEE (2016)

\bibitem{rauhofer}
Rauhofer, J.: ``{Privacy is dead, get over it!'' Information privacy and the
  dream of a risk-free society}. Information \& Communications Technology Law
  \textbf{17}(3),  185--197 (2008). \doi{10.1080/13600830802472990}

\bibitem{schwaber2004agile}
Schwaber, K.: Agile project management with Scrum. Microsoft press (2004)

\bibitem{sion2020neverending}
Sion, L., Landuyt, D.V., Joosen, W.: The never-ending story: On the need for
  continuous privacy impact assessment. In: 2020 IEEE European Symposium on
  Security and Privacy Workshops (EuroS PW). pp. 314--317. IEEE (2020).
  \doi{10.1109/EuroSPW51379.2020.00049}

\bibitem{sion2021automated}
Sion, L., Van~Landuyt, D., Yskout, K., Verreydt, S., Joosen, W.: Automated
  threat analysis and management in a continuous integration pipeline. 2021
  IEEE Secure Development (SecDev)  (2021)

\bibitem{spiekermann2012challenges}
Spiekermann, S.: The challenges of privacy by design. Communications of the ACM
   \textbf{55}(7),  38--40 (2012)

\bibitem{spiekermannEngineering}
Spiekermann, S., Cranor, L.F.: Engineering privacy. IEEE Transactions on
  Software Engineering  \textbf{35}(1),  67--82 (2009).
  \doi{10.1109/TSE.2008.88}

\bibitem{spiekermann2019inside}
Spiekermann, S., Korunovska, J., Langheinrich, M.: Inside the organization: Why
  privacy and security engineering is a challenge for engineers. Proceedings of
  the IEEE  \textbf{107}(3),  600--615 (2019). \doi{10.1109/JPROC.2018.2866769}

\bibitem{srivastava2018review}
Srivastava, P., Khan, R.: A review paper on cloud computing. International
  Journal of Advanced Research in Computer Science and Software Engineering
  \textbf{8}(6),  17--20 (2018)

\bibitem{stahl2017continuous}
Stahl, D., Martensson, T., Bosch, J.: Continuous practices and devops: beyond
  the buzz, what does it all mean? In: 2017 43rd Euromicro Conference on
  Software Engineering and Advanced Applications (SEAA). pp. 440--448. IEEE
  (2017)

\bibitem{sweeney2002k}
Sweeney, L.: k-anonymity: A model for protecting privacy. International Journal
  of Uncertainty, Fuzziness and Knowledge-Based Systems  \textbf{10}(05),
  557--570 (2002)

\bibitem{tai2016continuous}
Tai, S.: Continuous, trustless, and fair: Changing priorities in services
  computing. In: European Conference on Service-Oriented and Cloud Computing.
  pp. 205--210. Springer (2016)

\bibitem{un1948udhr}
{United Nations General Assembly}: {Universal Declaration of Human Rights
  {(UDHR)}} (1948)

\bibitem{Voigt2017}
Voigt, P., von~dem Bussche, A.: {Enforcement and Fines Under the GDPR}, pp.
  201--217. Springer International Publishing, Cham (2017)

\bibitem{voigt2017eu}
Voigt, P., Von~dem Bussche, A.: The {EU} general data protection regulation
  ({GDPR}). A Practical Guide, 1st Ed., Cham: Springer International Publishing
   \textbf{10},  3152676 (2017)

\bibitem{theRightToPrivacy}
Warren, S.D., Brandeis, L.D.: The right to privacy. Harvard Law Review
  \textbf{4}(5),  193--220 (1890). \doi{https://doi.org/10.2307/1321160}

\bibitem{whitman2003two}
Whitman, J.Q.: The two western cultures of privacy: Dignity versus liberty.
  Yale LJ  \textbf{113}, ~1151 (2003)

\bibitem{yarlagadda2021devops}
Yarlagadda, R.T.: {DevOps and Its Practices}. International Journal of Creative
  Research Thoughts (IJCRT), ISSN pp. 2320--2882 (2021)

\bibitem{zhou}
Zhou, M., Zhang, R., Xie, W., Qian, W., Zhou, A.: Security and privacy in cloud
  computing: A survey. In: 2010 Sixth International Conference on Semantics,
  Knowledge and Grids. pp. 105--112 (2010). \doi{10.1109/SKG.2010.19}

\bibitem{zimmermann2020}
Zimmermann, C.: Automation potentials in privacy engineering. In: Roßnagel,
  H., Schunck, C.H., Mödersheim, S., Hühnlein, D. (eds.) Open Identity Summit
  2020. pp. 121--132. Gesellschaft für Informatik e.V., Bonn (2020).
  \doi{10.18420/ois2020\_10}

\bibitem{zuboff2019age}
Zuboff, S.: The age of surveillance capitalism: The fight for a human future at
  the new frontier of power. Profile books (2019)

\end{thebibliography}

\end{document}